\documentclass[twocolumn]{aastex631}

\shorttitle{Eruption of a magnetic flux rope in a Radiative MHD Simulation}
\shortauthors{Chen, Rempel \& Fan}

\usepackage{natbib}
\usepackage{amsmath}
\usepackage{graphicx,subfigure}
\usepackage[percent]{overpic}
\usepackage{color}
\hypersetup{linkcolor=red,citecolor=blue}

\newcommand{\sectref}[1]{Section\,\ref{#1}}

\newcommand{\figref}[1]{Figure\,\ref{#1}}

\begin{document}
\nolinenumbers

\title{Eruption of a Magnetic Flux Rope in a Comprehensive \\Radiative Magnetohydrodynamic Simulation of flare-productive active regions}

\author[0000-0002-1963-5319]{Feng Chen}
\affiliation{School of Astronomy and Space Science, Nanjing University, Nanjing, 210023, China}
\affiliation{Key Laboratory of Modern Astronomy and Astrophysics (Nanjing University), Ministry of Education, Nanjing, 210023, China}
\affiliation{Laboratory for Atmospheric and Space Physics, University of Colorado, Boulder, 80303, USA}
\affiliation{High Altitude Observatory, National Center for Atmospheric Research, Boulder, 80307, USA}
\author[0000-0001-5850-3119]{Matthias Rempel}
\affiliation{High Altitude Observatory, National Center for Atmospheric Research, Boulder, 80307, USA}
\author[0000-0003-1027-0795]{Yuhong Fan}
\affiliation{High Altitude Observatory, National Center for Atmospheric Research, Boulder, 80307, USA}

\correspondingauthor{Feng Chen}
\email{chenfeng@nju.edu.cn}

\begin{abstract}
Radiative magnetohydrodynamic simulation includes sufficiently realistic physics to allow for the synthesis of remote sensing observables that can be quantitatively compared with observations. We analyze the largest flare in a simulation of the emergence of large flare-productive active regions described by Chen et al. The flare releases $4.5\times10^{31}$\,erg of magnetic energy and is accompanied by a spectacular coronal mass ejection. Synthetic soft X-ray flux of this flare reaches M2 class. The eruption reproduces many key features of observed solar eruptions. A pre-existing magnetic flux rope is formed along the highly sheared polarity inversion line between a sunspot pair and is covered by an overlying multi-pole magnetic field. During the eruption, the progenitor flux rope actively reconnects with the canopy field and evolves to the large-scale multi-thermal flux rope that is observed in the corona. Meanwhile, the magnetic energy released via reconnection is channeled down to the lower atmosphere and gives rise to bright soft X-ray post-flare loops and flare ribbons that reproduce the morphology and dynamic evolution of observed flares. The model helps to shed light on questions of where and when the a flux rope may form and how the magnetic structures in an eruption are related to observable emission properties.
\end{abstract}


\section{INTRODUCTION}\label{sec:intro}
A general picture of solar flares as a result of abrupt release of magnetic energy via reconnection has been well summarized by classical two-dimensional models \citep{Priest+Forbes:2002,Shibata+Magara:2011}.  A three-dimensional standard flares model proposed by \citet{Aulanier+al:2012} connects observed signatures of flares with the topological properties of the reconnected magnetic field. Since decades ago, flares and coronal mass ejections (CMEs) have been considered in a unified scenario of solar eruptions \citep[e.g.,][]{Shibata+al:1995}. CME models pay more attention to the trigger of eruptions and propagation of the erupted magnetic structure \citep{Forbes+Isenberg:1991,Antiochos+al:1999,Lin+Forbes:2000,Chen+Shibata:2000,Moore+al:2001,Toeroek+al:2004}. The abundant modeling works on CMEs and their connection with observations in the past several decades have been comprehensively reviewed and discussed by \citet{ChenPF:2011,Janvier+al:2015,Cheng+al:2017,Green+al:2018,Archontis+Syntelis:2019,Georgoulis+al:2019}. 

Realistic models of solar eruptions that bring together the sophisticated energetic consideration of flaring plasma as in one-/two-dimensional models \citep{Fisher+al:1985,Kowalski:2017} and the 3D geometry and magnetic topology similar to actual eruptions \citep[e.g.,][]{Amari+al:2003,Aulanier+al:2010,Kusano+al:2012,Wyper+al:2017,Jiang+al:2021} are immensely needed to bridge the gap between models and observations. With a focus on the evolution of an initially unstable eruption progenitor in the large-scale corona, some MHD models have taken into account a more realistic treatment of plasma thermodynamics \citep{JinMeng+al:2017,Downs+al:2021,Fan:2022} and allowed model synthesized observables to be compared with observations. \citet{Cheung+al:2019} reported a 3D radiative MHD simulation that accounts for optically thick radiative transport in the photosphere. Such treatment allows a more self-consistent description of flux emergence starting from the convective zone and meaningful synthesis of observables in the lower solar atmosphere. 

With the same numerical method, \citet{Chen+al:2022} simulated the evolution of flare-productive active regions by the emergence of active-region-scale flux bundles. Over a time period of 48 hours, more than one hundred flares that are larger than B-class occur spontaneously as an integral part of the emergence of active regions. Following the preceding paper that describes the general properties of the quiet Sun and active region evolution, we present in this paper an overview of the strongest and most spectacular flare event in the simulation. The simulation setup is briefly reviewed in \sectref{sec:method}. \sectref{sec:res_cme} shows the emission and magnetic properties of the event of interest. We compare the simulation results with observations in \sectref{sec:dis}.

\section{SIMULATION SETUP AND THE FLARE}\label{sec:method}
The simulation is performed with the MURaM code \citep{Rempel:2017}, which accounts for the radiative transfer under local thermal equilibrium approximation and optically thin radiation and field-aligned thermal conduction in the transition region and corona. The simulation domain spans 196.608\,Mm in the horizontal direction and 122.88\,Mm in the vertical direction, which is resolved by 1024 and 1920 grid points, yielding resolutions of 192 and 64\,km, respectively. The bottom boundary is placed at about 10\,Mm beneath the photosphere, where flux bundles generated in a solar convective dynamo \citep{Fan+Fang:2014} emerge into the domain as implemented by \citet{Chen+al:2017}. A detailed description of the simulation setup is presented in \citet{Chen+al:2022}. After an evolution of 48 hours solar time, the domain contains more than $10^{23}$\,Mx unsigned magnetic flux and close to $10^{33}$\,erg free magnetic energy, as well as complex shaped sunspots similar to flare productive active regions of the real Sun.

The event of interest, which is largest and arguably the most intriguing flare in the simulation, is initiated at $t{\approx}27$\,h42\,m and peaks at $t{=}27$\,h\,47\,m\,20\,s. The synthetic GOES flux of this flare reaches an absolute peak value of  $2.7\times10^{-5}$W m$^{-2}$ and a background deducted peak of $1.8\times10^{-5}$W m$^{-2}$. About $4.5\times10^{31}$\,erg of magnetic energy is released during the event, a large portion of which contributes to the kinetic energy of an erupted flux rope that eventually develops into a CME. The CME makes this flare an exceptional case in the entire simulation, as all the other eruptions that occur in the simulation are confined.

\section{RESULTS}\label{sec:res_cme}
\begin{figure*}
\center
\gridline{\fig{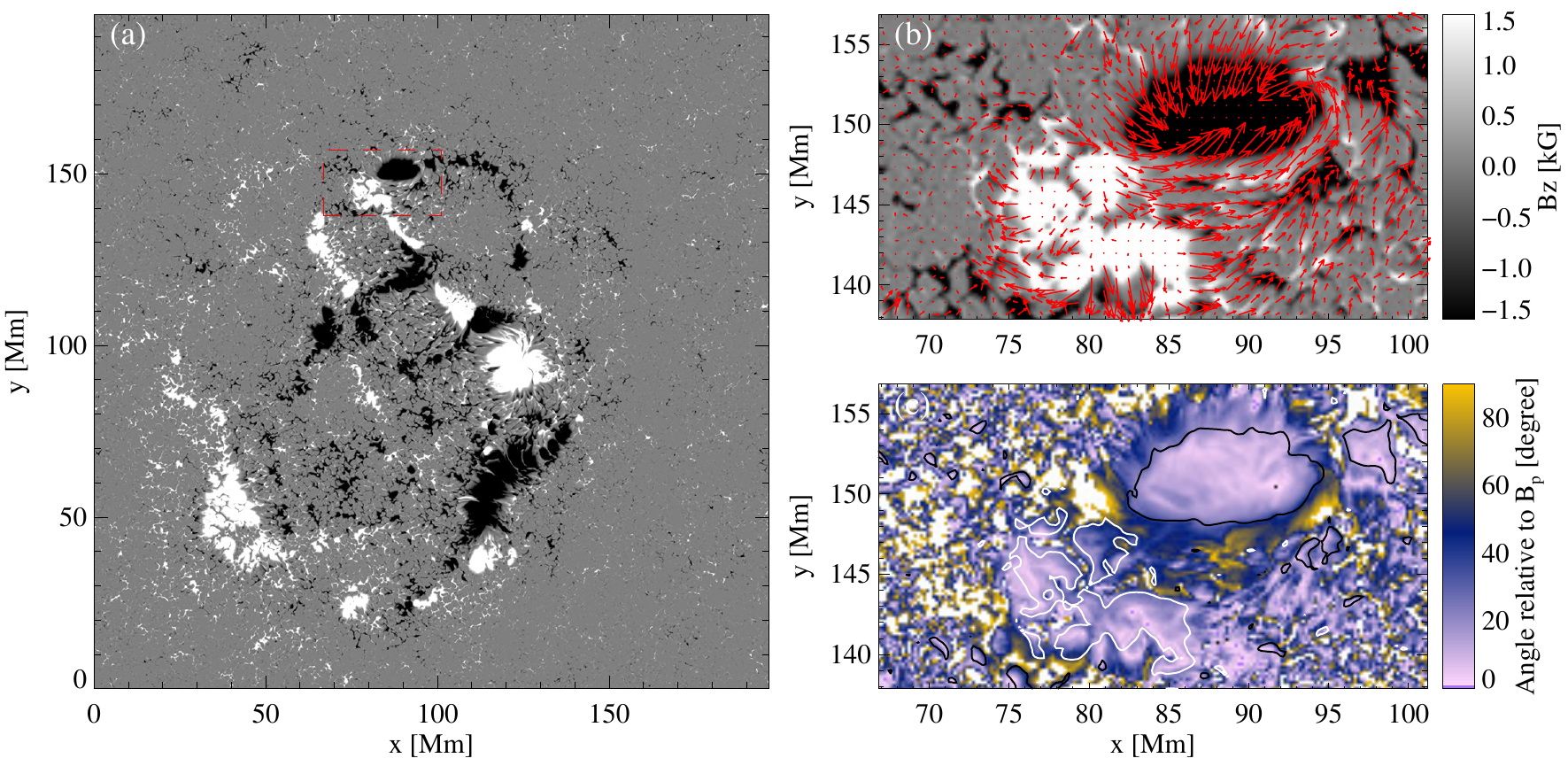}{\textwidth}{}}
\gridline{\fig{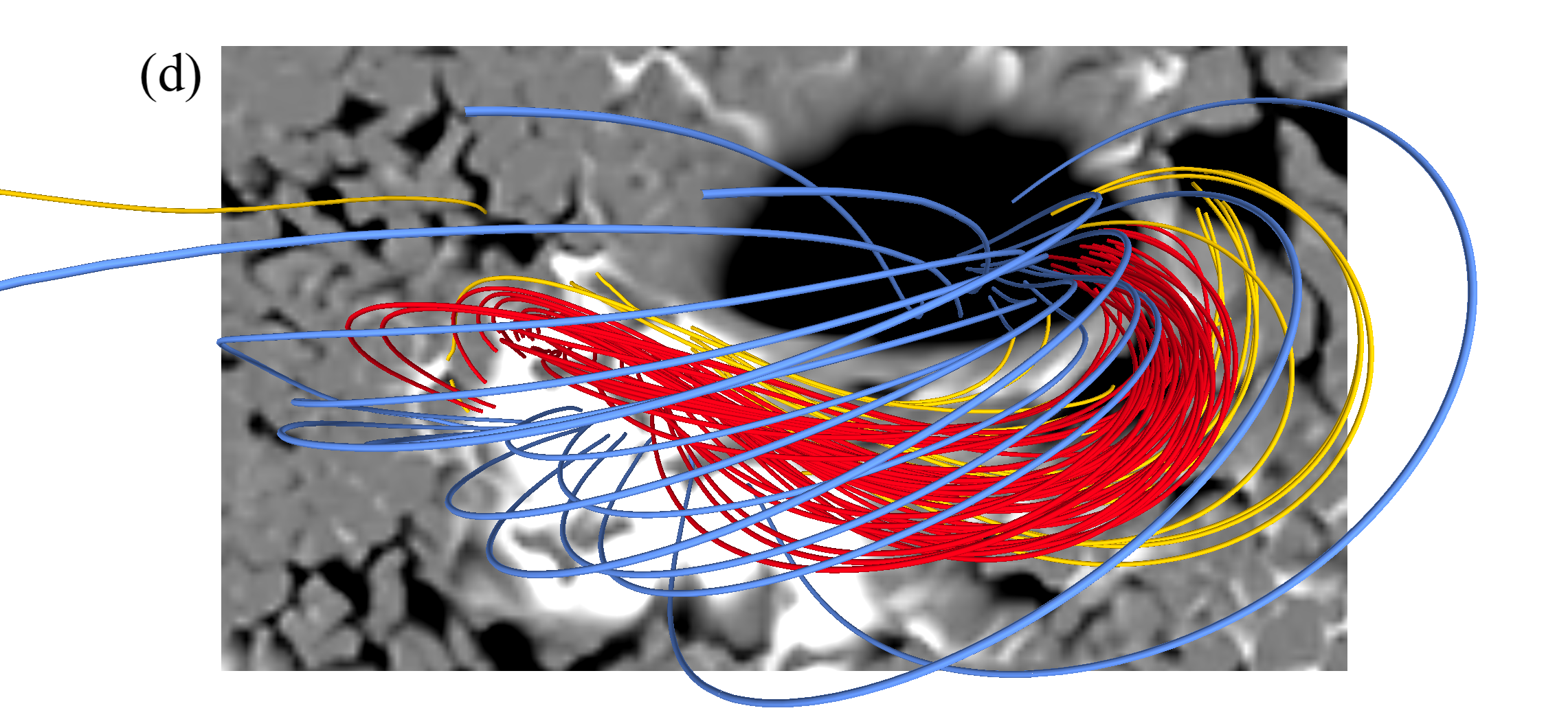}{0.5\textwidth}{}\fig{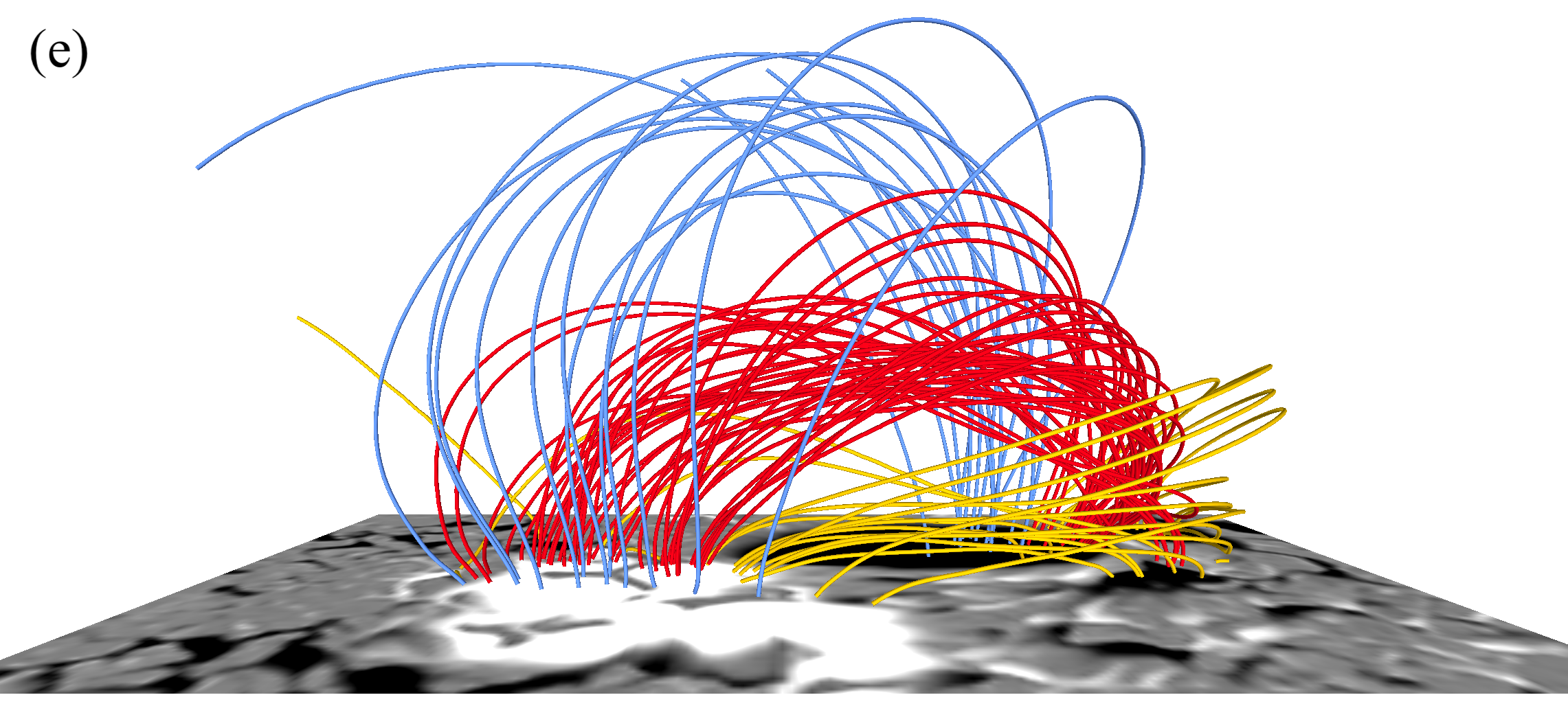}{0.5\textwidth}{}}
\caption{Magnetic field of the source region of the eruption. (a): Line-of-sight magnetic field ($B_{z}$) in the photosphere (surface of optical depth of unity). The red box indicates the source active region of the flare. (b): $B_{z}$ in the source region. Arrows show the vector of the horizontal magnetic field. (c): Angles between the actual magnetic field (full vector) and the potential field in the photosphere. (d): Magnetic field lines above the PIL. Field lines are calculated from seed points that are randomly distributed above the PIL. Seed points for the blue lines spread between $z=9.4$ and 14 Mm, and seed points for the red lines are in height range between $z=2.8$ and 7.5 Mm. Golden lines are calculated from seed points distributed between $z=1.0$ and 2.5 Mm. (e): The same field lines from a horizontal view.
\label{fig:ar_cme_mag}}
\end{figure*}

\begin{figure*}
\center
\includegraphics{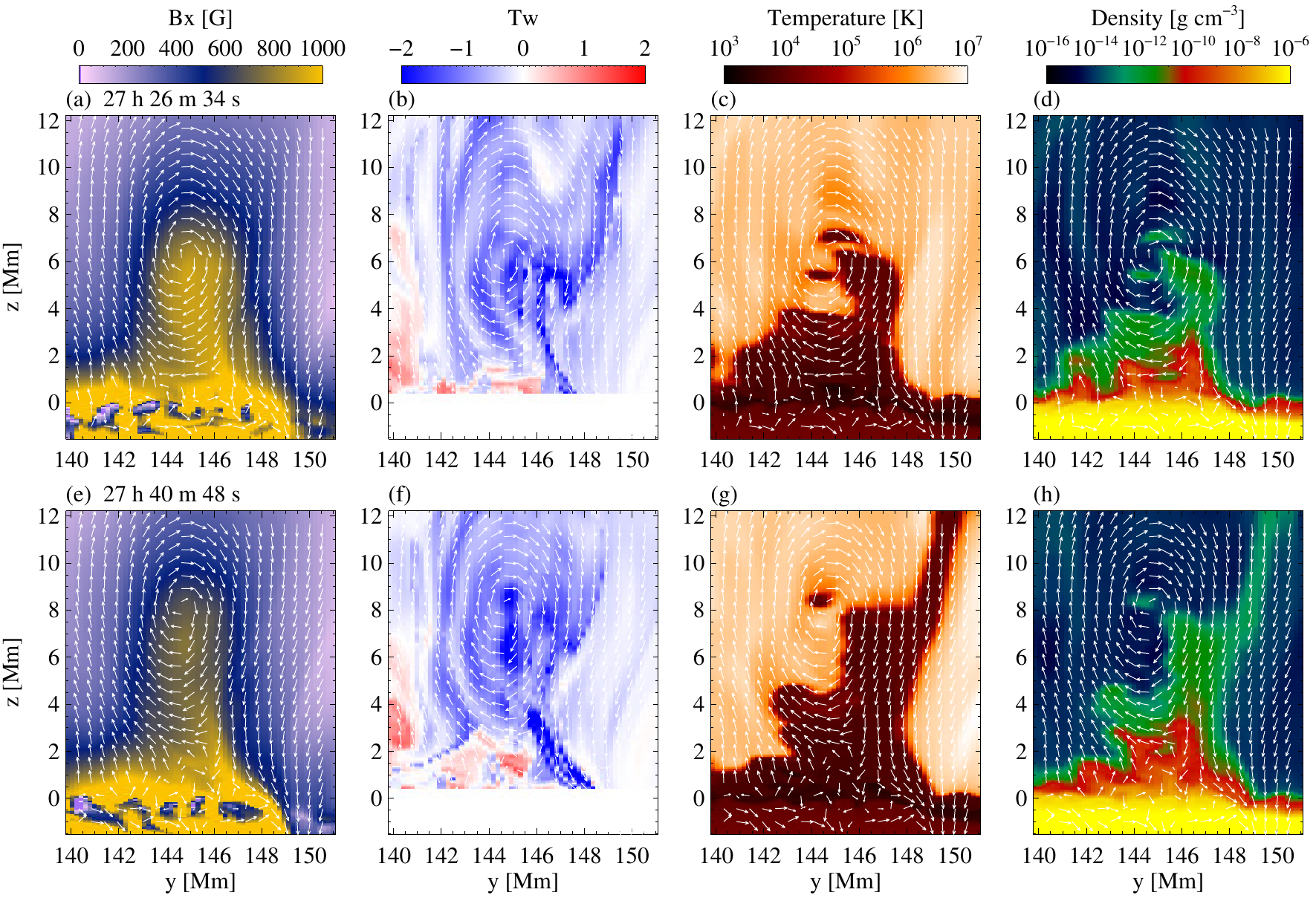}
\caption{Magnetic field and plasma thermal properties in a vertical cut ($y-z$ plane) at $x{=}87$\,Mm. (a): $B_{x}$. Arrows show the direction of the magnetic field in the plane ($B_{y}$ and $B_{z}$). (b): Magnetic twist integrated along field lines (evaluated for each grid point above $z=0.5$ km). (c) \& (d): Plasma temperature and density in the same vertical cut. Panels (a)--(d) and (e)--(f) present the same quantities for two snapshots, respectively.}
\label{fig:ar_cme_subpho}
\end{figure*}

\subsection{Source Region of the Eruption}
The flare occurs in the bipole at the north (larger $y$) edge of the complex active region formed by the emerged flux bundles, as shown by the photospheric ling-of-sight magnetic field in \figref{fig:ar_cme_mag}(a). \figref{fig:ar_cme_mag}(b) shows a zoomed view of the sunspot pair, with the arrows showing the horizontal magnetic field\footnote{Arrows are only plotted in the region of $|B_{z}|\leq2000$\,G to improve the visibility.}. It is clear that the horizontal magnetic field between the two spots is strongly sheared and apparently wraps around the negative polarity sunspot. To further illustrate the strong deviation of the magnetic field from the potential field, we evaluate the angle $\theta_{p}$ between the actual magnetic field and the potential field by
\begin{equation}
\theta_{\rm p} = \arccos\left(\frac{\mathbf{B}\cdot\mathbf{B_{p}}}{|\mathbf{B}||\mathbf{B_{p}}|}\right),
\end{equation}
where $\mathbf{B}$ is the magnetic vector on the surface of optical depth ($\tau$) of unity, and the potential field $B_{\rm p}$ is calculated from the $B_{z}$ in the same $\tau$ surface\footnote{This option mimics the effect in observations that measures the magnetic field in a tau surface instead of a geometrically horizontal layer. We have compared the results by using the magnetic field in a horizontal slice that corresponds to the photospheric height. The fact that the horizontal magnetic field is strongly sheared along the PIL and wraps the sunspots is not affected.}, with the condition of vanishing $B_{z}$ at infinity and periodic side boundaries (for the entire domain). The angle is presented in \figref{fig:ar_cme_mag}(c) with a color scale saturated at $90^{\circ}$ and contour lines of $B_{z}{=}\pm1500$\,G that outline the sunspots. The magnetic field inside the sunspots is close to a potential field. It is particularly interesting to note that spots are surrounded by a highly non-potential field, as indicated by angles near or larger than 90$^{\circ}$, while the region right in between the two spots is of smaller angles. This suggests a non-uniform distribution of shear along the PIL, whose implication will be discussed in combination with the observational properties of the flare. \figref{fig:ar_cme_mag}(d) \& (e) present the magnetic field lines above the PIL. The red lines manifest a magnetic flux rope with an axis roughly following the sheared PIL. The gold lines illustrate the shear arcades blow the flux rope. The right (larger $x$) end of the low-lying arcade wraps the negative sunspot and roots mostly in the two small flux concentrations. The magnetic field above the flux rope is, as shown by the blue lines, more aligned with the potential field (i.e., directly connecting the sunspots, or namely, normal to the PIL).

In \figref{fig:ar_cme_subpho}, we present the magnetic and plasma properties in a $y-z$ plane that intersects the flux rope at $x{=}87$\,Mm. The upper and lower rows correspond to two snapshots that are 14\,min apart. Panel (a) shows $B_{x}$ that mostly represents the axial field of the flux rope, while the arrows clearly illustrate the twist component. The axis of the flux rope rises slowly and steadily for about 2 Mm during this time period, yielding an average rising speed of 2.4 km s$^{-1}$. \figref{fig:ar_cme_subpho}(b) displays the magnetic twist $T_{w}$ evaluated by 
\begin{equation}\label{equ:twist}
T_{w}=\frac{1}{4\pi}\int_{L}\frac{(\nabla\times\mathbf{B})\cdot\mathbf{B}}{\mathbf{B}^2}dl,
\end{equation}
where the integral is done along the magnetic field line passing through each grid point ($z\ge$0.5 Mm) in this slice. $T_{w}$ measures the twist of an individual field line and is suggested to be a reasonable approximation of the twist number of a cylindrical flux tube near its axis \citep{LiuRui+al:2016}. The actual flux rope is a more complex structure than an ideal cylindrical flux tube. Nevertheless, the $T_{w}$ map shows that the flux rope is in general weakly twisted (less than two turns), which is similar to the findings in a number of previous numerical studies \citep[as reviewed by][]{Patsourakos+al:2020}. Moreover, the twist distribution is highly nonuniform, and the area around the axis of the flux rope have higher $T_{w}$, which also slightly increases with time, as demonstrated by the increased area of highly twisted field lines ($T_{w}>2$) in \figref{fig:ar_cme_subpho}(f). A similar twist distribution and evolution was also found by \citet{LiuRui+al:2016} in observations. \figref{fig:ar_cme_subpho} (c) \& (d) and (g) \& (h) show the chromospheric plasma that is trapped in the flux rope and carried upward as the magnetic structure rises. This plasma is the major source of the emission features during the impulsive eruption of the flux rope presented in the following section.

\subsection{Observable Features of the Eruption}

\begin{figure*}
\center
\includegraphics{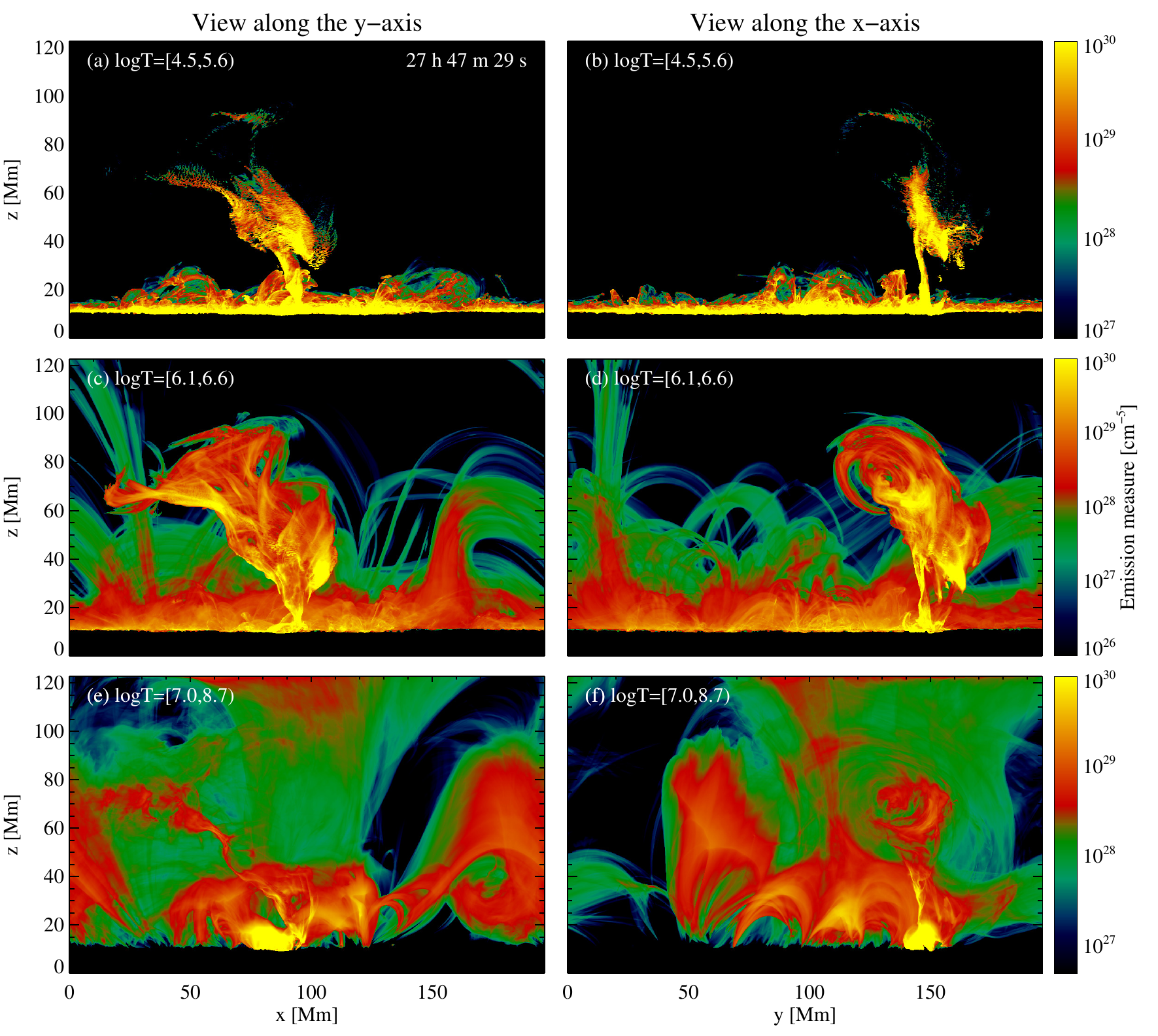}
\caption{Emission measures in three temperature ranges. The left column shows a line-of-sight along the $y$-axis of the domain. The right column shows a view along the $x$-axis of the domain, corresponding to observations on the east limb of the Sun. The time stamp of this snapshot is 9\,s after the flare peak. An animation that covers the dynamics evolution of 39\,min starting from $t{=}27$\,h\,28\,m\,2\,s (about 20\,min before the flare peak) is available in the electronic version. 
\label{fig:ar_cme_em}}
\end{figure*}

\begin{figure*}
\center
\includegraphics{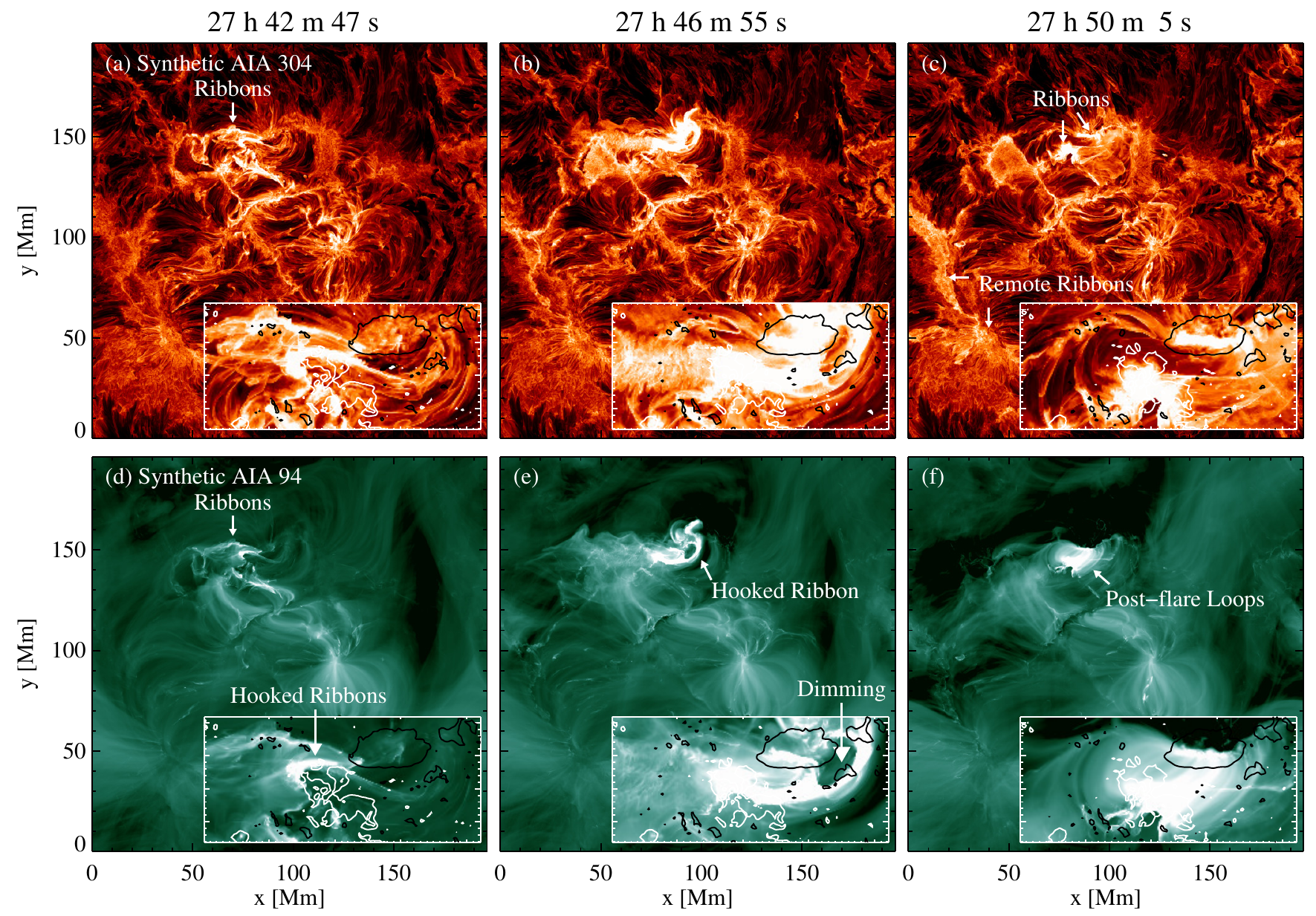}
\caption{Synthetic AIA 304\,\AA~and 94\,\AA~channel images from the viewpoint at the top of the domain. Large panels show the whole horizontal domain, while the inserts present a zoomed view of the source region of the eruption. Contour lines in the inserts mark $B_{z}{=}\pm1500$\,G.  Arrows indicate features discussed in the text. An animation that covers the dynamics evolution of 39\,min starting from $t{=}27$\,h\,28\,m\,2\,s (about 20\,min before the flare peak) is available in the electronic version.
\label{fig:ar_cme_ribbon}}
\end{figure*}

\subsubsection{General Evolution of The Flux Rope Eruption}
The emission measure (EM) is a key high-cadence output of the simulation \citep{Rempel:2017,Chen+al:2022} and serves as the primary tool for diagnosing plasma thermodynamics as done in observations. We present in \figref{fig:ar_cme_em} the EMs integrated over three temperature ranges. This snapshot is 9\,s after the GOES flux peak, with an animated version covering the full course of the eruption.

The animation shows that the very first signature of the eruption can be seen in the frame near $t{=}27$\,h\,41\,m\,30\,s in the low-temperature EM in panel (b) at $y{\approx}150$\,Mm and in the high-temperature EM in panel (e) at $x{\approx}65$\,Mm, which means that the eruption is initiated at one end of the flux rope, instead of the middle part of it. The asymmetric eruption quickly develops into a cone-like structure that expands and rotates while rising upward, as displayed by the middle-temperature EM in \figref{fig:ar_cme_em}(c). The mass load in the initial flux rope is mostly in chromospheric temperatures ($~10^{4}$\,K), as shown in \figref{fig:ar_cme_subpho}(d). However, the erupting flux rope is mostly magnificently seen transition-region and coronal temperatures, which indicate a significant heating of the trapped plasma during the eruption.

The high-temperature EM along the $x$-axis in panel (f) shows that the erupting flux rope encounters some canopy loops at $z{\approx}60$\,Mm ($t{\approx}27$\,h\,43\,m in the animation). The flux rope pushes them upward and evolves into a new structure by reconnection with the overlying magnetic field, as presented later in the paper. The location where reconnection occurs under the erupting flux rope is outlined in \figref{fig:ar_cme_em}(e) by a thin inclined thread of hot plasma (from $(x,z){=}(100,10)$\,Mm to $(x,z){=}(50,60)$\,Mm), which is roughly along the low edge of the cooler cone-like structure shown in panel (c). The brightest feature in the high-temperature EM is a series of post-flare loops found around $x{\approx}80$\,Mm and $y{=}150$\,Mm. These loops are the main emitter in the soft X-ray passbands and give rise to the GOES flux peak.

\subsubsection{Flare Ribbons and Dimmings}
We present in \figref{fig:ar_cme_ribbon} synthetic AIA images in the 304\,\AA~\footnote{Although the interpretation of observations of the 304\,\AA~channel  needs to consider the possible blending of coronal emission lines, the synthetic images are based on a temperature response function that captures chromosphere and transition region plasma.}~and 94\,\AA channels from a vertical view. In the three snapshots shown in \figref{fig:ar_cme_ribbon}, we highlight a few features in the simulated eruption, which are also commonly seen in observations. An animated version of the figure covers the full course of the eruption.

The first snapshot (left column of \figref{fig:ar_cme_ribbon}) captures the very beginning phase of the flare. The insets in each panel shows a zoomed view of the source region with contour lines of $B_{z}{=}\pm1500$\,G. A hook-shaped bright ribbon can be seen in both 304\,\AA~and 94\,\AA~images, as indicated by the arrows.  The ribbon is found to wrap around the positive polarity spot. Meanwhile, a companion ribbon appears almost simultaneously  in the negative polarity magnetic-element region north of the spot. The evolution shown by the animation demonstrates that the hooked ribbon first appears on the left edge of the positive spot and propagates rightward along the boundary ($B_{z}$ contour) of the spot. This trend is continued in the next evolution stage as presented below.

The second snapshot (middle column of \figref{fig:ar_cme_ribbon}) captures the later development of the flare ribbons in the negative polarity spot region. The right end of the ribbon also appears in a hooked shape and reaches two small flux concentrations next to the negative spot, which are the foot points of the arcade that wraps the right leg the flux rope before the eruption, as shown in \figref{fig:ar_cme_mag}. Meanwhile, we see a formation of a counterpart ribbon at the southern (smaller $y$) edge of the negative spot. The evolution of the bright ribbons, which is initiated near $x{\approx}60$\,Mm and propagates through the PIL, maps the development of reconnection in the coronal magnetic field. This evolution is also consistent with that of the EM along the $y-axis$, as the asymmetric eruption of the flux rope is led by its left end, and the post-flare loops form following the same trend. 

A dimming region, which is observed more clearly in the 94\,\AA~image, is formed in the region enclosed by the hooked ribbon and the sunspot. A comparison with \figref{fig:ar_cme_mag} and \figref{fig:ar_cme_em} suggests that this dimming region is where the highly sheared horizontal magnetic field anchors and where the apex of the cone-like structure, i.e., one of the foot points of the flux rope, is rooted.

The third snapshot (right column of \figref{fig:ar_cme_ribbon}) highlights the main (post peak) phase of the flare. The 304\,\AA~image shown in \figref{fig:ar_cme_ribbon}(c) provides a clear view of the flare ribbons. We find that the ribbon in the negative spot concentrates at the boundary of the coherent spot and that the ribbon in the positive spot appears as a patch that covers the area of the discrete spot. The 94\,\AA~image shows the same flare ribbons, as well as the hot loops connecting them. The temporal evolution shows that the two flare ribbons continuously expand outward for a few minutes after the snapshot shown here, which implies that more magnetic flux is reconnected. It is also interesting to note that new ``remote ribbons" are formed on a much larger scale at locations indicated by the arrows in \figref{fig:ar_cme_ribbon}(c). The remote ribbons brighten as they extend from $y{\approx}100\,Mm$ downward to $y{\approx}40$\,Mm. This propagation maps the progress of reconnection between the erupting flux rope and the large-scale canopy magnetic field, as shown in the following section.

\subsection{Magnetic Field Involved in the Eruption}
\begin{figure*}
\center
\includegraphics{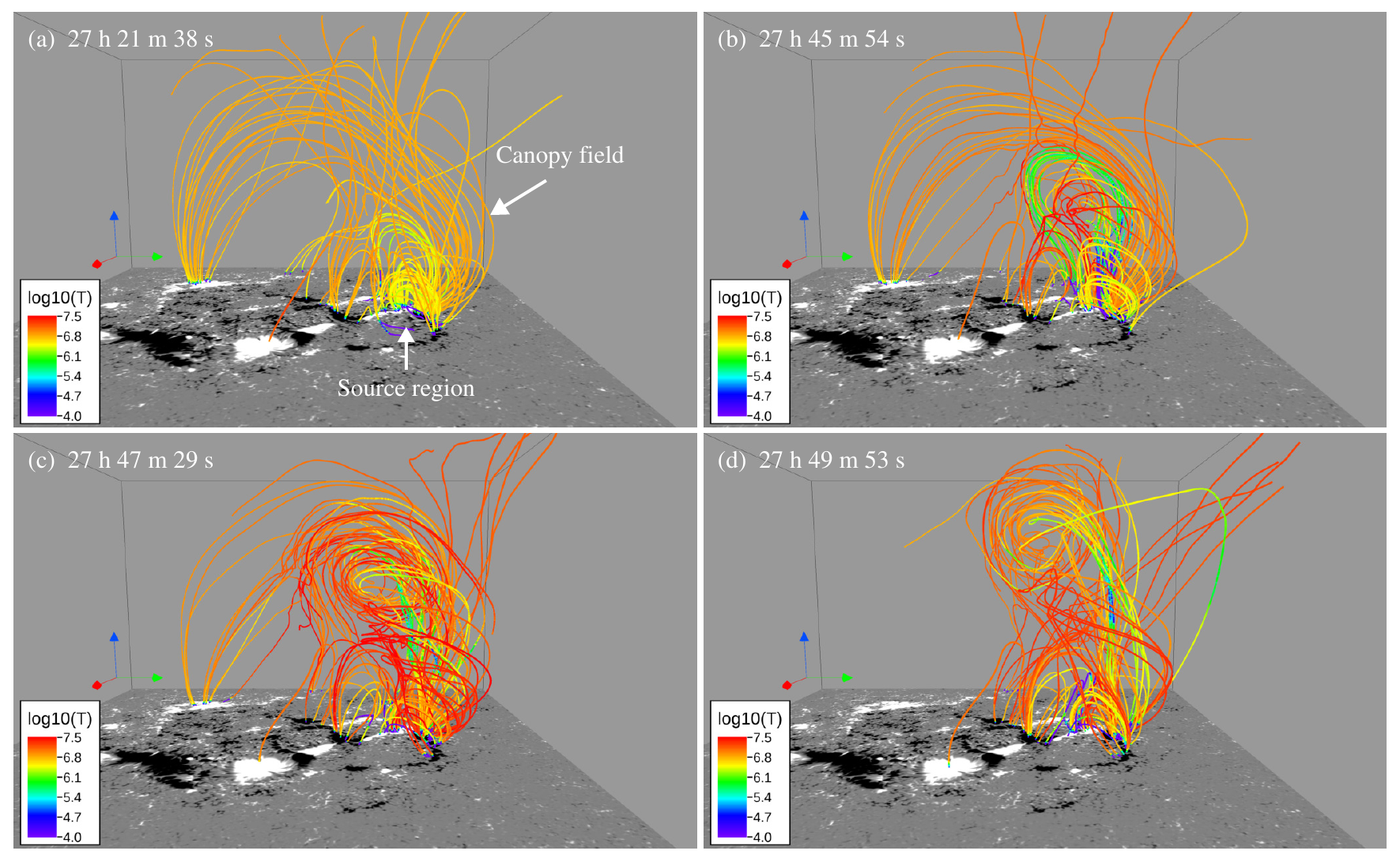}
\caption{Large-scale magnetic field before and during the eruption. The gray-scale images display $B_{z}$ at the photosphere.  Red, green, \& blue arrows indicate the $x$, $y$, and $z$ directions, respectively. Colors of the field lines represent plasma temperatures. (a): About 26\,min before the flare peak representing the pre-eruption state. (b): Impulsive phase of the flare (about 1.5\,min before the flare peak. (c): 9\,s seconds after the flare peak. (d):  2.5\,min after the peak time. An animated version of this figure that displays the evolution of 28\,min between Panels (a) and (d) is available in the electronic version.
\label{fig:ar_cme_large}} 
\end{figure*}

\begin{figure*}
\center
\includegraphics{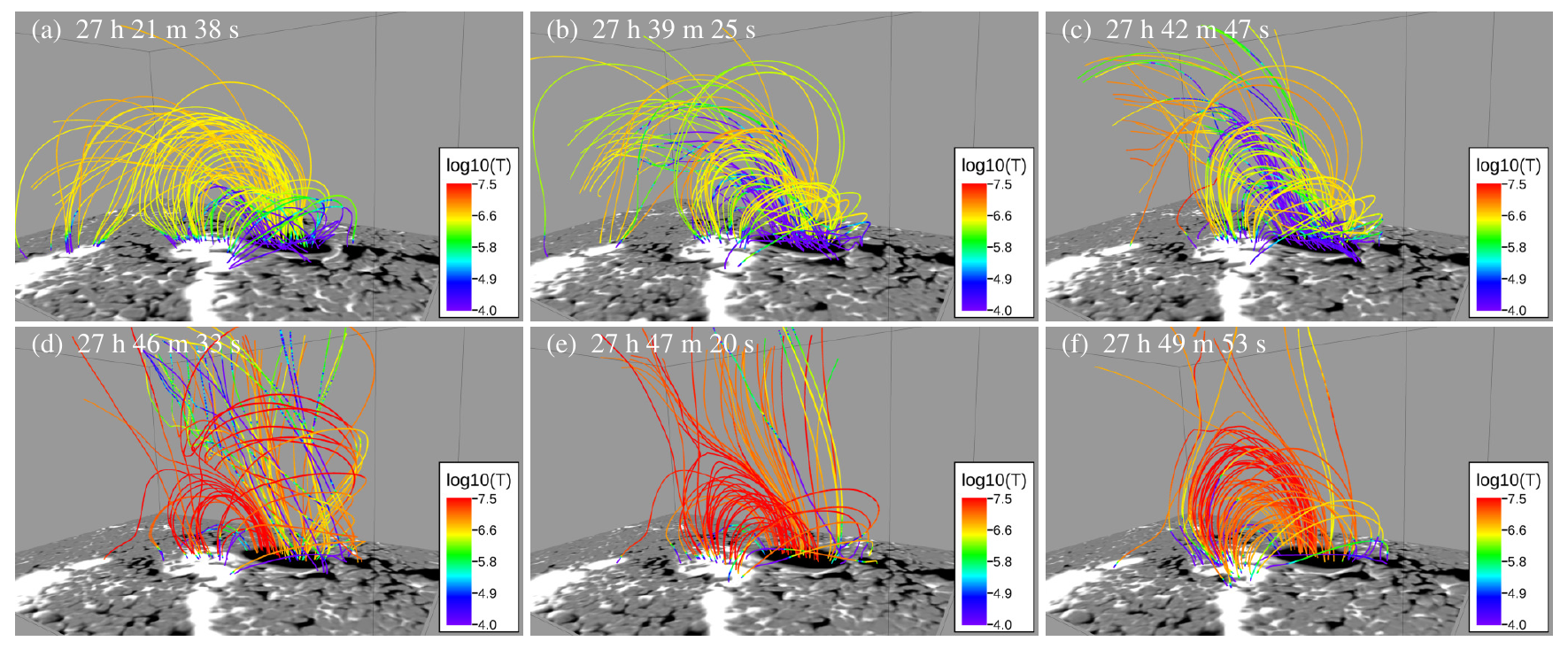}
\caption{Magnetic field lines above the source region of the eruption. Colors of the field lines represent plasma temperatures. The frame only indicates the boundary of the volume cropped from the whole simulation domain for this visualization. An animated version of this figure that displays the evolution of 28\,min between Panels (a) and (f) is available in the electronic version.
\label{fig:ar_cme_small}} 
\end{figure*}

\subsubsection{Large-scale Flux Rope}
In \figref{fig:ar_cme_large}, we display the large-scale coronal magnetic field involved in the eruption at four timestamps. The field lines are calculated from seed points, which are randomly distributed in the corona ($z>15$\,Mm) above the region spanning from $x{\approx}50$ to 110\,Mm and from $y{\approx}110$ to 150\,Mm. The lines are color coded according to plasma temperatures. An animated version of \figref{fig:ar_cme_large} displays the evolution of 28\,min covered by this figure.

Panel (a) of \figref{fig:ar_cme_large} shows the coronal magnetic field about 20\,min before the flare peak. The most evident magnetic structures are some open field lines connected to a negative spot near the center of the domain and a large-scale magnetic arch connecting the negative spot in the source region of the eruption and positive magnetic concentrations close to the corner of the domain (indicated as ``canopy field"). The flux rope is not evident in this large-scale view because the distribution of the seed points is not optimized for capturing it. Nevertheless, it can be found next to the foot point the canopy field as a bundle of field lines hosting cool plasma.

Panel (b) of \figref{fig:ar_cme_large} captures the magnetic field configuration in the impulsive phase. At this moment, the flux rope erupts and carries cool material into the corona, as indicated by the purple-colored lines, particularly around the axis of the flux rope. The erupting flux rope encounters the overlying arch and pushes the canopy upward. Reconnection between the flux rope and ambient magnetic field is illustrated by red-colored field lines, where the plasma is heated to well over $10^{7}$\,K.

Panel (c) \& (d) of \figref{fig:ar_cme_large} provide the best view of the large-scale flux rope at the flare peak. In particular, Panel (c) shows the same snapshot and a similar view angle as \figref{fig:ar_cme_em}(d). The temperature structure of the flux rope, which is illustrated by the colors of the field lines  is consistent with the multi-thermal EM shown in \figref{fig:ar_cme_em}: the cool axis is surrounded by million K plasma, while the hottest plasma (over 10 MK) wraps at the outermost shell with apparently the largest twist. The hot shell is a consequence of reconnection between the erupted flux rope and the ambient magnetic field, i.e., the open field and large-scale canopy field seen in Panels (a) and (b). The reconnection is also manifested by the fact that some of the field lines constituting the shell connect to the negative spot near the center of the domain (foot point of the open field before the eruption) and the positive spots near the corner of the domain (one foot point of the canopy field). The far side positive sunspot and flux concentrations are co-spatial with the remote flare ribbons shown in \figref{fig:ar_cme_ribbon}(c), which also suggests that they are foot points of reconnected field lines.

It is interesting to note that, for the case here, the large-scale flux rope, which is clearly visible and distinguished only when it reaches a sufficient height and size, is an outcome of the expansion of the initial flux rope in the source region and the reconnection of the initial flux rope with the ambient magnetic field. The latter contribution plays a vital role in creating the magnetic and thermal structure of the flux rope that are eventually observed on large spatial scales. The evolution from a small and weakly twisted flux rope to a large and highly twisted flux rope was also found in a numerical study by \citet{Syntelis+al:2017}. Recently, \citet{GouTingyu+al:2023} reported an observation of the formation of a large-scale flux rope and dynamical evolution of flare ribbons as the result of the reconnection between a seed flux rope and ambient magnetic field.

\subsubsection{Reconnection under the Erupted Flux Rope}
In \figref{fig:ar_cme_small}, we zoom into the source region of the initial flux rope. Similar to \figref{fig:ar_cme_large}, field lines are calculated from seed points that are fixed in time and randomly distributed over the sunspot pair to capture both flux rope along the PIL and overlying magnetic field lines connecting the two sunspots. Lines are color coded by plasma temperatures as done for \figref{fig:ar_cme_large}. An animated version of \figref{fig:ar_cme_small} covers the evolution of 28\,min.

Panels (a) - (c) of \figref{fig:ar_cme_small} show the magnetic field over this sunspot pair from more than 20\,min before the flare peak to the begin of the impulsive phase. The flux rope lies low along the PIL and is filled with plasma cooler than $10^{5}$\,K. A quasi-static rise of the flux rope, which is shown in \figref{fig:ar_cme_subpho}, is illustrated here by the slow rise of the cold axis and the upward expansion of the field lines connecting the two sunspots (more clearly shown by the animation). It also becomes clear that the field lines around the foot point of the initial flux rope make a hooked shape and wrap around the negative spot. This is a key structure that gives rise to the hooked bright ribbon shown in the middle column of \figref{fig:ar_cme_ribbon}.

Panel (d) of \figref{fig:ar_cme_small} shows a moment in the impulsive phase, which lies between panels (b) and (c) of \figref{fig:ar_cme_large}. The flux rope has erupted and propagated out of the field-of-view for this visualization, leaving only one of its legs, which is highly stretched upright. During this stage, the leg of the flux rope become almost anti-parallel to the magnetic field arcade that connect to the positive spot. This facilitates reconnection between them that forms cusp-shaped loops under the flux rope and converts the arcade to a component around the flux rope leg, as illustrated by the red-colored (hotter than $10^{7.5}$ MK) field lines in \figref{fig:ar_cme_small}(d).

Panels (e) and (f) of \figref{fig:ar_cme_small} present a clear view of the further development of the magnetic reconnection under the erupted flux rope. The field lines connecting the two spots are stretched upward and pushed toward each other. Reconnection is triggered when anti-parallel magnetic fields encounter each other. An elongated current sheet is formed in the layer between the anti-polarity field lines, which is the bright hot thread connecting the erupted flux rope above and the cusp-shaped loops below, as shown in \figref{fig:ar_cme_em}(f). A large amount of magnetic energy released by the reconnection is channeled down by thermal conduction to the chromosphere, where evaporation of plasma that is heated to tens of MK fills the cusp-shaped post-flare loops and gives rise to the high-temperature observable emission features displayed in  \figref{fig:ar_cme_em} and \figref{fig:ar_cme_ribbon}. The snapshot shown in \figref{fig:ar_cme_small}(d) is close to the earliest time ($t{\approx}27$\,h\,45\,m) when post-flare loops can be seen in the animation of \figref{fig:ar_cme_em}, and \figref{fig:ar_cme_small}(e) shows the snapshot at the flare peak time, when the post-flare loops contribute the strongest soft X-ray emission.

\section{Summary and Discussion}\label{sec:dis}
We present the second part in a series of studies on a radiative MHD simulation of solar active regions, following the first part that presents the general properties of the quiet Sun and non-eruptive stage of active regions \citet{Chen+al:2022}.  In this study, we focus on the magnetic and emission properties of the largest flare. The results are summarized and discussed as follows. 

The flare originates from the eruption of a flux rope along a strongly sheared PIL in a bipole active region. The eruption demonstrates a highly asymmetric shape as one non-active foot point keeps being anchored next to the sunspot, while the other active foot point whips upward. The erupted progenitor flux rope evolves in reconnection with the overlying canopy magnetic field, forming a large scale cone-shaped flux rope. When viewed from the side, the eruption shows a configuration of a multi-thermal flux rope,  thin and hot current sheet, and cusp-shaped post-flare loops, which is consistent with classical models and observations. Flare ribbons are generated by a fully self-consistent evolution of plasma thermodynamics properties in response to the disposition of energy in the lower atmosphere. 

\subsection{Pre-eruption Magnetic Structures}
Strongly sheared magnetic fields along PILs (sometimes referred simply as sheared PILs), which imply a strong non-potential magnetic field, have been found for decades \citep[e.g.,][]{Heyvaerts+al:1977,SunXudong+al:2012}. Whether the pre-eruption magnetic structure is a flux rope or sheared arcades remains a question \citep{Patsourakos+al:2020}. Observations have suggested that flux ropes can form from less twisted configurations (e.g., sheared arcades) on various time scales prior to the eruption \citep[e.g.,][]{WangHaimin:2015,James+al:2017,Chintzoglou+al:2019}. Shortly before the eruption, we see in \figref{fig:ar_cme_mag}(d) \& (e) a flux rope (red lines) with a prolonged axis above the photospheric PIL, while field lines immediately under the flux rope are shorter sheared arcades (golden lines) that appear in a hooked shape around the sunspots (clearer on the negative side). This configuration implies formation/growth of the flux rope via reconnection between the shorter sheared arcades, which has been found in previous numerical simulations \citep[see][and references therein]{Fan:2009,Leake+al:2013,Syntelis+al:2017,Toriumi+Takasao:2017} and also deduced from observable emission features such as formation of a sigmoid from two J-shaped structures \citep[see e.g., ][]{Green+al:2011,Savcheva+al:2012}.

The overlying magnetic field also plays a non-trivial role in solar eruptions. Although the whole domain demonstrates a very complex magnetogram, as shown in \figref{fig:ar_cme_mag}, the overall large-scale configuration of the magnetic field that is most relevant to this eruption is similar to that considered in the ``break out" model \citet{Antiochos+al:1999} and that observed in limb active regions \citet[e.g.,][]{Mason+al:2021}. The ``break out" model suggests that the reconnection of ambient magnetic fields opens up the confinement above sheared arcades in the source region. A similar process is also seen in this simulation, as implied by the fact that hot plasma appears in the overlying canopy field (see e.g., $(x, z)=(130,50)$ Mm at $t{\approx}$27\,h\,40\,m in Panel (f) of the animation of \figref{fig:ar_cme_em}) before the impulsive eruption of the flux rope. It is also interesting to note that the pre-eruption structure in our simulation is a magnetic flux rope, which may also be destabilized by, for example, torus instability \citep{Kliem+Toeroek:2006}. How the instability criteria deduced for more idealized setups (e.g., straight or symmetric torus-shaped tubes) can be applied in the complex magnetic configuration in this eruption requires a further investigation. Nonetheless, we suspect that the two processes may likely be cooperative in triggering and driving the eruption, which needs to be substantiated by further analysis of the data.

\subsection{Observational Signatures of Flux Ropes}
Asymmetric eruptions were also reported by, for example, \citet{LiuRui+al:2009,Joshi+al:2013,LiuChang+al:2015}. By comparing EUV observations and a numerical model of \citet{Shiota+al:2005}, \citet{Tripathi+al:2006,Li+Zhang:2009} interpreted the one-directional propagation of bright post-flare loops, which is similar to that shown in \figref{fig:ar_cme_ribbon}, as an indication of sequential reconnection of the ambient magnetic field triggered by an asymmetric eruption of a flux rope. 

When viewed along the $x$-axis (see \figref{fig:ar_cme_em}), the asymmetric geometry of the eruption collapses into a picture that is more similar to classical two-dimensional models \citep[e.g.,][]{Shibata+al:1995} and observations of limb flares triggered by flux rope eruption \citep[e.g.,][]{ChenBin+al:2020}. The plasma that is confined in and affected by the flux rope during the eruption spreads over the whole temperature of the simulation, leading to a multi-thermal flux rope. In observations, although a high-temperature structure was proposed as a proxy of a flux rope \citep[e.g.,][]{Zhang+al:2012}, it is now commonly found that flux ropes demonstrate DEMs over a broad temperature range \citep[e.g.,][ and references therein]{ChengXin+al:2012,Hannah+Kontar:2013,LiLeping+al:2022}. In particular, we notice a remarkable similarity between the  low-temperature EM shown in \figref{fig:ar_cme_em}(b) and the AIA 304\,\AA~channel image of flux rope observed on 2011 March 7th \citep[see Fig. 4 of ][]{ChengXin+al:2012}\footnote{The view point of the right column \figref{fig:ar_cme_em} (from positive to negative $x$ ) corresponds to observing the structure on the east limb, whereas the observation of \citet{ChengXin+al:2012} was on the west limb.}

The interaction between an erupting flux rope and overlying magnetic fields is supposed to be universal, although the overlying field may not always be illustrated by EUV loops. In the early phase of the eruption, the overlying field is pushed upward by the flux rope, similar to the process found in observations \citep{ChengXin+al:2011}. In the later evolution, the asymmetric flux rope undergoes reconnection with the large-scale overlying canopy field, and the active foot point switches to farther flux concentrations, which marks a clear difference between this eruption and the symmetric configuration considered in most models. In another experiment with the open top boundary\footnote{Top boundary is switched to open before the eruption, which changes the magnetic field in the high part of the domain due to increased plasma outflows, while the low-lying progenitor flux rope is not affected.}, the eruption is a more symmetric, indicating that the appearance of an eruption is very sensitive to how the erupting flux rope interacts with the overlying field.

\subsection{Observational Signatures of Flares: Hooked Ribbons and Dimmings}
Bright ribbons are a key observable property of solar flares. The flare ribbons of this eruption resemble the hooked flare ribbons that are commonly observed in solar flares. Combining observations and MHD or nonlinear forcefree modeling of the coronal magnetic field, such flare ribbons are interpreted as the imprint of quasi-separatrix layers of the coronal magnetic field where reconnection occurs in the 3D stand flare model  \citep[e.g.,][]{Janvier+al:2014,Inoue+al:2014,Savcheva+al:2015,ZhaoJie+al:2016}. These studies omitted a more realistic treatment for the plasma evolution. In our simulation, flare ribbons are generated from a self-consistent evolution of plasma thermodynamic properties in response to the magnetic energy that is released during reconnection and channeled down by a strong thermal conductive flux. This approach allows a direct connection between the magnetic topology in the corona and the observable flare ribbons and hence consolidates the previous results that are obtained primarily based on magnetic modeling.

Dimming regions embraced by hooked flare ribbons are assumed to be the foot points of erupted flux ropes and hence are used to estimate the axial flux of flux ropes  \citep[e.g.,][]{ChengXin+Ding:2016,WangWensi+al:2017,XingChen+al:2020}. This assumption can be validated in this simulation, as we see that the dimming corresponds to the root of the flux rope that wraps the sunspot. More specifically, ``flux rope" here refers to the large-scale flux rope formed through reconnection between the initial flux rope and the ambient field. This is demonstrated by the fact that the hooked flare ribbon seen in \figref{fig:ar_cme_ribbon}(e) is cospatial with the two negative flux concentrations, which are the foot points of the arcade that wraps the leg of the initial flux rope as shown in \figref{fig:ar_cme_mag}, and that this arcade is reconnected with the erupted flux rope as shown in \figref{fig:ar_cme_small}(d). This circumstance is consistent with that assumed when analyzing observational data because the large-scale flux rope corresponds to the observed emission structures.

Moreover, the fast extension of flare ribbons along both sides of the PIL and the following separation between the flare ribbons reproduce the ribbon evolution commonly observed in flares. Observations by \citet{ChengXin+Ding:2016} and \citet{Qiu+Cheng:2022} noticed that flare ribbons tend to be strongly sheared at the two ends but weakly sheared in the middle. This is consistent with the non-uniform distribution of the shear angle of the magnetic field along the PIL in our simulation.

\subsection{Concluding Remark}
The M2 flare is the strongest flare achieved thus far in radiative MHD simulations that are capable of giving quantitative estimates of flare classes.  Although the simulation is not set up to model any particular active region or eruption event, the dynamo-generated magnetic and flow fields provide sufficient realism in the sunspots formed in the photosphere and coronal magnetic field. Therefore, this flare is able to reproduce many key general magnetic and emission features observed in real solar eruptions.

By coupling with a global-scale convective dynamo simulation, this model will help us to trace the origin of solar eruptions back to the solar interior, where the active region magnetic flux is generated and address formation of the flux rope above the photosphere is related to the sub-surface magnetic and flow fields in the context of flux emergence. Furthermore, the pre-eruption magnetic field exhibits a combination of  a pre-existing flux rope within ``break out" like canopy field. This raises intriguing questions of whether reconnection or instability acts as the trigger and how these can be differentiated from the observable emission properties.

\begin{acknowledgements}
We thank the referee for comments that improve the clarity of this letter. F.C. is supported by the National Key R\&D Program of China under grant 2021YFA1600504 and the Program for Innovative Talents and Entrepreneurs in Jiangsu. This material is based upon work supported by the National Center for Atmospheric Research, which is a major facility sponsored by the National Science Foundation under Cooperative Agreement No. 1852977. We would like to acknowledge high-performance computing support from Cheyenne (doi:10.5065/D6RX99HX) provided by NCAR’s Computational and Information Systems Laboratory, sponsored by the National Science Foundation. F.C. had been supported by the Advanced Study Program postdoctoral fellowship at NCAR and by the George Ellery Hale postdoctoral fellowship at the University of Colorado Boulder since this project was initiated in 2016. The 3D visualization in \figref{fig:ar_cme_mag}, \figref{fig:ar_cme_large}, and \figref{fig:ar_cme_small} are produced by VAPOR \citep{VAPOR}.
\end{acknowledgements}

\bibliography{reference}

\begin{thebibliography}{}
\expandafter\ifx\csname natexlab\endcsname\relax\def\natexlab#1{#1}\fi
\providecommand{\url}[1]{\href{#1}{#1}}
\providecommand{\dodoi}[1]{doi:~\href{http://doi.org/#1}{\nolinkurl{#1}}}
\providecommand{\doeprint}[1]{\href{http://ascl.net/#1}{\nolinkurl{http://ascl.net/#1}}}
\providecommand{\doarXiv}[1]{\href{https://arxiv.org/abs/#1}{\nolinkurl{https://arxiv.org/abs/#1}}}

\bibitem[{{Amari} {et~al.}(2003){Amari}, {Luciani}, {Aly}, {Mikic}, \&
  {Linker}}]{Amari+al:2003}
{Amari}, T., {Luciani}, J.~F., {Aly}, J.~J., {Mikic}, Z., \& {Linker}, J. 2003,
  \apj, 585, 1073, \dodoi{10.1086/345501}

\bibitem[{{Antiochos} {et~al.}(1999){Antiochos}, {DeVore}, \&
  {Klimchuk}}]{Antiochos+al:1999}
{Antiochos}, S.~K., {DeVore}, C.~R., \& {Klimchuk}, J.~A. 1999, \apj, 510, 485,
  \dodoi{10.1086/306563}

\bibitem[{{Archontis} \& {Syntelis}(2019)}]{Archontis+Syntelis:2019}
{Archontis}, V., \& {Syntelis}, P. 2019, Philosophical Transactions of the
  Royal Society of London Series A, 377, 20180387,
  \dodoi{10.1098/rsta.2018.0387}

\bibitem[{{Aulanier} {et~al.}(2012){Aulanier}, {Janvier}, \&
  {Schmieder}}]{Aulanier+al:2012}
{Aulanier}, G., {Janvier}, M., \& {Schmieder}, B. 2012, \aap, 543, A110,
  \dodoi{10.1051/0004-6361/201219311}

\bibitem[{{Aulanier} {et~al.}(2010){Aulanier}, {T{\"o}r{\"o}k}, {D{\'e}moulin},
  \& {DeLuca}}]{Aulanier+al:2010}
{Aulanier}, G., {T{\"o}r{\"o}k}, T., {D{\'e}moulin}, P., \& {DeLuca}, E.~E.
  2010, \apj, 708, 314, \dodoi{10.1088/0004-637X/708/1/314}

\bibitem[{{Chen} {et~al.}(2020){Chen}, {Yu}, {Reeves}, \&
  {Gary}}]{ChenBin+al:2020}
{Chen}, B., {Yu}, S., {Reeves}, K.~K., \& {Gary}, D.~E. 2020, \apjl, 895, L50,
  \dodoi{10.3847/2041-8213/ab901a10.48550/arXiv.2005.01900}

\bibitem[{{Chen} {et~al.}(2017){Chen}, {Rempel}, \& {Fan}}]{Chen+al:2017}
{Chen}, F., {Rempel}, M., \& {Fan}, Y. 2017, \apj, 846, 149,
  \dodoi{10.3847/1538-4357/aa85a0}

\bibitem[{{Chen} {et~al.}(2022){Chen}, {Rempel}, \& {Fan}}]{Chen+al:2022}
---. 2022, \apj, 937, 91, \dodoi{10.3847/1538-4357/ac8f95}

\bibitem[{{Chen}(2011)}]{ChenPF:2011}
{Chen}, P.~F. 2011, Living Reviews in Solar Physics, 8, 1,
  \dodoi{10.12942/lrsp-2011-1}

\bibitem[{{Chen} \& {Shibata}(2000)}]{Chen+Shibata:2000}
{Chen}, P.~F., \& {Shibata}, K. 2000, \apj, 545, 524, \dodoi{10.1086/317803}

\bibitem[{{Cheng} \& {Ding}(2016)}]{ChengXin+Ding:2016}
{Cheng}, X., \& {Ding}, M.~D. 2016, \apjs, 225, 16,
  \dodoi{10.3847/0067-0049/225/1/1610.48550/arXiv.1605.04047}

\bibitem[{{Cheng} {et~al.}(2017){Cheng}, {Guo}, \& {Ding}}]{Cheng+al:2017}
{Cheng}, X., {Guo}, Y., \& {Ding}, M. 2017, Science China Earth Sciences, 60,
  1383, \dodoi{10.1007/s11430-017-9074-6}

\bibitem[{{Cheng} {et~al.}(2011){Cheng}, {Zhang}, {Liu}, \&
  {Ding}}]{ChengXin+al:2011}
{Cheng}, X., {Zhang}, J., {Liu}, Y., \& {Ding}, M.~D. 2011, \apjl, 732, L25,
  \dodoi{10.1088/2041-8205/732/2/L2510.48550/arXiv.1103.5084}

\bibitem[{{Cheng} {et~al.}(2012){Cheng}, {Zhang}, {Saar}, \&
  {Ding}}]{ChengXin+al:2012}
{Cheng}, X., {Zhang}, J., {Saar}, S.~H., \& {Ding}, M.~D. 2012, \apj, 761, 62,
  \dodoi{10.1088/0004-637X/761/1/6210.48550/arXiv.1210.7287}

\bibitem[{{Cheung} {et~al.}(2019){Cheung}, {Rempel}, {Chintzoglou}, {Chen},
  {Testa}, {Mart{\'\i}nez-Sykora}, {Sainz Dalda}, {DeRosa}, {Malanushenko},
  {Hansteen}, {De Pontieu}, {Carlsson}, {Gudiksen}, \&
  {McIntosh}}]{Cheung+al:2019}
{Cheung}, M.~C.~M., {Rempel}, M., {Chintzoglou}, G., {et~al.} 2019, Nature
  Astronomy, 3, 160, \dodoi{10.1038/s41550-018-0629-3}

\bibitem[{{Chintzoglou} {et~al.}(2019){Chintzoglou}, {Zhang}, {Cheung}, \&
  {Kazachenko}}]{Chintzoglou+al:2019}
{Chintzoglou}, G., {Zhang}, J., {Cheung}, M. C.~M., \& {Kazachenko}, M. 2019,
  \apj, 871, 67, \dodoi{10.3847/1538-4357/aaef30}

\bibitem[{{Downs} {et~al.}(2021){Downs}, {Warmuth}, {Long}, {Bloomfield},
  {Kwon}, {Veronig}, {Vourlidas}, \& {Vr{\v{s}}nak}}]{Downs+al:2021}
{Downs}, C., {Warmuth}, A., {Long}, D.~M., {et~al.} 2021, \apj, 911, 118,
  \dodoi{10.3847/1538-4357/abea78}

\bibitem[{{Fan}(2009)}]{Fan:2009}
{Fan}, Y. 2009, \apj, 697, 1529, \dodoi{10.1088/0004-637X/697/2/1529}

\bibitem[{{Fan}(2022)}]{Fan:2022}
---. 2022, \apj, 941, 61, \dodoi{10.3847/1538-4357/aca0ec}

\bibitem[{{Fan} \& {Fang}(2014)}]{Fan+Fang:2014}
{Fan}, Y., \& {Fang}, F. 2014, \apj, 789, 35,
  \dodoi{10.1088/0004-637X/789/1/35}

\bibitem[{{Fisher} {et~al.}(1985){Fisher}, {Canfield}, \&
  {McClymont}}]{Fisher+al:1985}
{Fisher}, G.~H., {Canfield}, R.~C., \& {McClymont}, A.~N. 1985, \apj, 289, 414,
  \dodoi{10.1086/162901}

\bibitem[{{Forbes} \& {Isenberg}(1991)}]{Forbes+Isenberg:1991}
{Forbes}, T.~G., \& {Isenberg}, P.~A. 1991, \apj, 373, 294,
  \dodoi{10.1086/170051}

\bibitem[{{Georgoulis} {et~al.}(2019){Georgoulis}, {Nindos}, \&
  {Zhang}}]{Georgoulis+al:2019}
{Georgoulis}, M.~K., {Nindos}, A., \& {Zhang}, H. 2019, Philosophical
  Transactions of the Royal Society of London Series A, 377, 20180094,
  \dodoi{10.1098/rsta.2018.0094}

\bibitem[{Gou {et~al.}(2023)Gou, Liu, Veronig, Zhuang, Li, Wang, Xu, \&
  Wang}]{GouTingyu+al:2023}
Gou, T., Liu, R., Veronig, A.~M., {et~al.} 2023, Nature Astronomy,
  \dodoi{10.1038/s41550-023-01966-2}

\bibitem[{{Green} {et~al.}(2011){Green}, {Kliem}, \& {Wallace}}]{Green+al:2011}
{Green}, L.~M., {Kliem}, B., \& {Wallace}, A.~J. 2011, \aap, 526, A2,
  \dodoi{10.1051/0004-6361/201015146}

\bibitem[{{Green} {et~al.}(2018){Green}, {T{\"o}r{\"o}k}, {Vr{\v{s}}nak},
  {Manchester}, \& {Veronig}}]{Green+al:2018}
{Green}, L.~M., {T{\"o}r{\"o}k}, T., {Vr{\v{s}}nak}, B., {Manchester}, W., \&
  {Veronig}, A. 2018, \ssr, 214, 46, \dodoi{10.1007/s11214-017-0462-5}

\bibitem[{{Hannah} \& {Kontar}(2013)}]{Hannah+Kontar:2013}
{Hannah}, I.~G., \& {Kontar}, E.~P. 2013, \aap, 553, A10,
  \dodoi{10.1051/0004-6361/201219727}

\bibitem[{{Heyvaerts} {et~al.}(1977){Heyvaerts}, {Priest}, \&
  {Rust}}]{Heyvaerts+al:1977}
{Heyvaerts}, J., {Priest}, E.~R., \& {Rust}, D.~M. 1977, \apj, 216, 123,
  \dodoi{10.1086/155453}

\bibitem[{{Inoue} {et~al.}(2014){Inoue}, {Hayashi}, {Magara}, {Choe}, \&
  {Park}}]{Inoue+al:2014}
{Inoue}, S., {Hayashi}, K., {Magara}, T., {Choe}, G.~S., \& {Park}, Y.~D. 2014,
  \apj, 788, 182, \dodoi{10.1088/0004-637X/788/2/182}

\bibitem[{{James} {et~al.}(2017){James}, {Green}, {Palmerio}, {Valori}, {Reid},
  {Baker}, {Brooks}, {van Driel-Gesztelyi}, \& {Kilpua}}]{James+al:2017}
{James}, A.~W., {Green}, L.~M., {Palmerio}, E., {et~al.} 2017, \solphys, 292,
  71, \dodoi{10.1007/s11207-017-1093-410.48550/arXiv.1703.10837}

\bibitem[{{Janvier} {et~al.}(2014){Janvier}, {Aulanier}, {Bommier},
  {Schmieder}, {D{\'e}moulin}, \& {Pariat}}]{Janvier+al:2014}
{Janvier}, M., {Aulanier}, G., {Bommier}, V., {et~al.} 2014, \apj, 788, 60,
  \dodoi{10.1088/0004-637X/788/1/6010.48550/arXiv.1402.2010}

\bibitem[{{Janvier} {et~al.}(2015){Janvier}, {Aulanier}, \&
  {D{\'e}moulin}}]{Janvier+al:2015}
{Janvier}, M., {Aulanier}, G., \& {D{\'e}moulin}, P. 2015, \solphys, 290, 3425,
  \dodoi{10.1007/s11207-015-0710-3}

\bibitem[{{Jiang} {et~al.}(2021){Jiang}, {Feng}, {Liu}, {Yan}, {Hu}, {Moore},
  {Duan}, {Cui}, {Zuo}, {Wang}, \& {Wei}}]{Jiang+al:2021}
{Jiang}, C., {Feng}, X., {Liu}, R., {et~al.} 2021, Nature Astronomy, 5, 1126,
  \dodoi{10.1038/s41550-021-01414-z}

\bibitem[{{Jin} {et~al.}(2017){Jin}, {Manchester}, {van der Holst}, {Sokolov},
  {T{\'o}th}, {Vourlidas}, {de Koning}, \& {Gombosi}}]{JinMeng+al:2017}
{Jin}, M., {Manchester}, W.~B., {van der Holst}, B., {et~al.} 2017, \apj, 834,
  172, \dodoi{10.3847/1538-4357/834/2/172}

\bibitem[{{Joshi} {et~al.}(2013){Joshi}, {Srivastava}, {Filippov}, {Uddin},
  {Kayshap}, \& {Chand ra}}]{Joshi+al:2013}
{Joshi}, N.~C., {Srivastava}, A.~K., {Filippov}, B., {et~al.} 2013, \apj, 771,
  65, \dodoi{10.1088/0004-637X/771/1/65}

\bibitem[{{Kliem} \& {T{\"o}r{\"o}k}(2006)}]{Kliem+Toeroek:2006}
{Kliem}, B., \& {T{\"o}r{\"o}k}, T. 2006, \prl, 96, 255002,
  \dodoi{10.1103/PhysRevLett.96.255002}

\bibitem[{{Kowalski} {et~al.}(2017){Kowalski}, {Allred}, {Daw}, {Cauzzi}, \&
  {Carlsson}}]{Kowalski:2017}
{Kowalski}, A.~F., {Allred}, J.~C., {Daw}, A., {Cauzzi}, G., \& {Carlsson}, M.
  2017, \apj, 836, 12, \dodoi{10.3847/1538-4357/836/1/12}

\bibitem[{{Kusano} {et~al.}(2012){Kusano}, {Bamba}, {Yamamoto}, {Iida},
  {Toriumi}, \& {Asai}}]{Kusano+al:2012}
{Kusano}, K., {Bamba}, Y., {Yamamoto}, T.~T., {et~al.} 2012, \apj, 760, 31,
  \dodoi{10.1088/0004-637X/760/1/31}

\bibitem[{{Leake} {et~al.}(2013){Leake}, {Linton}, \&
  {T{\"o}r{\"o}k}}]{Leake+al:2013}
{Leake}, J.~E., {Linton}, M.~G., \& {T{\"o}r{\"o}k}, T. 2013, \apj, 778, 99,
  \dodoi{10.1088/0004-637X/778/2/99}

\bibitem[{{Li} {et~al.}(2022){Li}, {Song}, {Peter}, \&
  {Chitta}}]{LiLeping+al:2022}
{Li}, L., {Song}, H., {Peter}, H., \& {Chitta}, L.~P. 2022, \apjl, 941, L1,
  \dodoi{10.3847/2041-8213/aca47b10.48550/arXiv.2211.11148}

\bibitem[{{Li} \& {Zhang}(2009)}]{Li+Zhang:2009}
{Li}, L., \& {Zhang}, J. 2009, \apj, 690, 347,
  \dodoi{10.1088/0004-637X/690/1/347}

\bibitem[{Li {et~al.}(2019)Li, Jaroszynski, Pearse, Orf, \& Clyne}]{VAPOR}
Li, S., Jaroszynski, S., Pearse, S., Orf, L., \& Clyne, J. 2019, Atmosphere,
  10, \dodoi{10.3390/atmos10090488}

\bibitem[{{Lin} \& {Forbes}(2000)}]{Lin+Forbes:2000}
{Lin}, J., \& {Forbes}, T.~G. 2000, \jgr, 105, 2375,
  \dodoi{10.1029/1999JA900477}

\bibitem[{{Liu} {et~al.}(2015){Liu}, {Deng}, {Liu}, {Lee}, {Pariat},
  {Wiegelmann}, {Liu}, {Kleint}, \& {Wang}}]{LiuChang+al:2015}
{Liu}, C., {Deng}, N., {Liu}, R., {et~al.} 2015, \apjl, 812, L19,
  \dodoi{10.1088/2041-8205/812/2/L19}

\bibitem[{{Liu} {et~al.}(2009){Liu}, {Alexander}, \&
  {Gilbert}}]{LiuRui+al:2009}
{Liu}, R., {Alexander}, D., \& {Gilbert}, H.~R. 2009, \apj, 691, 1079,
  \dodoi{10.1088/0004-637X/691/2/1079}

\bibitem[{{Liu} {et~al.}(2016){Liu}, {Kliem}, {Titov}, {Chen}, {Wang}, {Wang},
  {Liu}, {Xu}, \& {Wiegelmann}}]{LiuRui+al:2016}
{Liu}, R., {Kliem}, B., {Titov}, V.~S., {et~al.} 2016, \apj, 818, 148,
  \dodoi{10.3847/0004-637X/818/2/148}

\bibitem[{{Mason} {et~al.}(2021){Mason}, {Antiochos}, \&
  {Vourlidas}}]{Mason+al:2021}
{Mason}, E.~I., {Antiochos}, S.~K., \& {Vourlidas}, A. 2021, \apjl, 914, L8,
  \dodoi{10.3847/2041-8213/ac0259}

\bibitem[{{Moore} {et~al.}(2001){Moore}, {Sterling}, {Hudson}, \&
  {Lemen}}]{Moore+al:2001}
{Moore}, R.~L., {Sterling}, A.~C., {Hudson}, H.~S., \& {Lemen}, J.~R. 2001,
  \apj, 552, 833, \dodoi{10.1086/320559}

\bibitem[{{Patsourakos} {et~al.}(2020){Patsourakos}, {Vourlidas},
  {T{\"o}r{\"o}k}, {Kliem}, {Antiochos}, {Archontis}, {Aulanier}, {Cheng},
  {Chintzoglou}, {Georgoulis}, {Green}, {Leake}, {Moore}, {Nindos}, {Syntelis},
  {Yardley}, {Yurchyshyn}, \& {Zhang}}]{Patsourakos+al:2020}
{Patsourakos}, S., {Vourlidas}, A., {T{\"o}r{\"o}k}, T., {et~al.} 2020, \ssr,
  216, 131, \dodoi{10.1007/s11214-020-00757-910.48550/arXiv.2010.10186}

\bibitem[{{Priest} \& {Forbes}(2002)}]{Priest+Forbes:2002}
{Priest}, E.~R., \& {Forbes}, T.~G. 2002, \aapr, 10, 313,
  \dodoi{10.1007/s001590100013}

\bibitem[{{Qiu} \& {Cheng}(2022)}]{Qiu+Cheng:2022}
{Qiu}, J., \& {Cheng}, J. 2022, \solphys, 297, 80,
  \dodoi{10.1007/s11207-022-02003-710.48550/arXiv.2205.03004}

\bibitem[{{Rempel}(2017)}]{Rempel:2017}
{Rempel}, M. 2017, \apj, 834, 10, \dodoi{10.3847/1538-4357/834/1/10}

\bibitem[{{Savcheva} {et~al.}(2015){Savcheva}, {Pariat}, {McKillop},
  {McCauley}, {Hanson}, {Su}, {Werner}, \& {DeLuca}}]{Savcheva+al:2015}
{Savcheva}, A., {Pariat}, E., {McKillop}, S., {et~al.} 2015, \apj, 810, 96,
  \dodoi{10.1088/0004-637X/810/2/96}

\bibitem[{{Savcheva} {et~al.}(2012){Savcheva}, {Green}, {van Ballegooijen}, \&
  {DeLuca}}]{Savcheva+al:2012}
{Savcheva}, A.~S., {Green}, L.~M., {van Ballegooijen}, A.~A., \& {DeLuca},
  E.~E. 2012, \apj, 759, 105, \dodoi{10.1088/0004-637X/759/2/105}

\bibitem[{{Shibata} \& {Magara}(2011)}]{Shibata+Magara:2011}
{Shibata}, K., \& {Magara}, T. 2011, Living Reviews in Solar Physics, 8, 6,
  \dodoi{10.12942/lrsp-2011-6}

\bibitem[{{Shibata} {et~al.}(1995){Shibata}, {Masuda}, {Shimojo}, {Hara},
  {Yokoyama}, {Tsuneta}, {Kosugi}, \& {Ogawara}}]{Shibata+al:1995}
{Shibata}, K., {Masuda}, S., {Shimojo}, M., {et~al.} 1995, \apjl, 451, L83,
  \dodoi{10.1086/309688}

\bibitem[{{Shiota} {et~al.}(2005){Shiota}, {Isobe}, {Chen}, {Yamamoto},
  {Sakajiri}, \& {Shibata}}]{Shiota+al:2005}
{Shiota}, D., {Isobe}, H., {Chen}, P.~F., {et~al.} 2005, \apj, 634, 663,
  \dodoi{10.1086/496943}

\bibitem[{{Sun} {et~al.}(2012){Sun}, {Hoeksema}, {Liu}, {Wiegelmann},
  {Hayashi}, {Chen}, \& {Thalmann}}]{SunXudong+al:2012}
{Sun}, X., {Hoeksema}, J.~T., {Liu}, Y., {et~al.} 2012, \apj, 748, 77,
  \dodoi{10.1088/0004-637X/748/2/77}

\bibitem[{{Syntelis} {et~al.}(2017){Syntelis}, {Archontis}, \&
  {Tsinganos}}]{Syntelis+al:2017}
{Syntelis}, P., {Archontis}, V., \& {Tsinganos}, K. 2017, \apj, 850, 95,
  \dodoi{10.3847/1538-4357/aa9612}

\bibitem[{{Toriumi} \& {Takasao}(2017)}]{Toriumi+Takasao:2017}
{Toriumi}, S., \& {Takasao}, S. 2017, \apj, 850, 39,
  \dodoi{10.3847/1538-4357/aa95c2}

\bibitem[{{T{\"o}r{\"o}k} {et~al.}(2004){T{\"o}r{\"o}k}, {Kliem}, \&
  {Titov}}]{Toeroek+al:2004}
{T{\"o}r{\"o}k}, T., {Kliem}, B., \& {Titov}, V.~S. 2004, \aap, 413, L27,
  \dodoi{10.1051/0004-6361:20031691}

\bibitem[{{Tripathi} {et~al.}(2006){Tripathi}, {Isobe}, \&
  {Mason}}]{Tripathi+al:2006}
{Tripathi}, D., {Isobe}, H., \& {Mason}, H.~E. 2006, \aap, 453, 1111,
  \dodoi{10.1051/0004-6361:20064993}

\bibitem[{{Wang} {et~al.}(2015){Wang}, {Cao}, {Liu}, {Xu}, {Liu}, {Zeng},
  {Chae}, \& {Ji}}]{WangHaimin:2015}
{Wang}, H., {Cao}, W., {Liu}, C., {et~al.} 2015, Nature Communications, 6,
  7008, \dodoi{10.1038/ncomms8008}

\bibitem[{{Wang} {et~al.}(2017){Wang}, {Liu}, {Wang}, {Hu}, {Shen}, {Jiang}, \&
  {Zhu}}]{WangWensi+al:2017}
{Wang}, W., {Liu}, R., {Wang}, Y., {et~al.} 2017, Nature Communications, 8,
  1330, \dodoi{10.1038/s41467-017-01207-x}

\bibitem[{{Wyper} {et~al.}(2017){Wyper}, {Antiochos}, \&
  {DeVore}}]{Wyper+al:2017}
{Wyper}, P.~F., {Antiochos}, S.~K., \& {DeVore}, C.~R. 2017, \nat, 544, 452,
  \dodoi{10.1038/nature22050}

\bibitem[{{Xing} {et~al.}(2020){Xing}, {Cheng}, \& {Ding}}]{XingChen+al:2020}
{Xing}, C., {Cheng}, X., \& {Ding}, M.~D. 2020, The Innovation, 1, 100059,
  \dodoi{10.1016/j.xinn.2020.10005910.48550/arXiv.2011.10750}

\bibitem[{{Zhang} {et~al.}(2012){Zhang}, {Cheng}, \& {Ding}}]{Zhang+al:2012}
{Zhang}, J., {Cheng}, X., \& {Ding}, M.-D. 2012, Nature Communications, 3, 747,
  \dodoi{10.1038/ncomms1753}

\bibitem[{{Zhao} {et~al.}(2016){Zhao}, {Gilchrist}, {Aulanier}, {Schmieder},
  {Pariat}, \& {Li}}]{ZhaoJie+al:2016}
{Zhao}, J., {Gilchrist}, S.~A., {Aulanier}, G., {et~al.} 2016, \apj, 823, 62,
  \dodoi{10.3847/0004-637X/823/1/62}

\end{thebibliography}
\bibliographystyle{aasjournal}

\end{document}